\RequirePackage{amsmath}
\documentclass{article}
\usepackage{graphicx}

\usepackage{tikz}
\usetikzlibrary{arrows,shapes,automata, calc, chains, matrix, trees, positioning, scopes}
\tikzset{snake it/.style={decorate, decoration=snake}}

\usepackage{amsthm,amsmath,amssymb,amsfonts}
\usepackage{bbm}
\usepackage{graphicx}
\usepackage{multirow}
\usepackage{xspace}
\usepackage[colorlinks=true,citecolor=blue,linkcolor=blue]{hyperref}
\usepackage{cleveref}
\usepackage{mathtools}

\usepackage{multicol}

\usepackage{enumerate}

\newtheorem{theorem}{Theorem}

\newtheorem{corollary}[theorem]{Corollary}

\usepackage[letterpaper,top=2cm,bottom=2cm,left=3cm,right=3cm,marginparwidth=1.75cm]{geometry}

\usepackage{fancyhdr}
\newcommand{\OO}{\mathcal{O}}
\newcommand{\Aa}{\mathcal{A}}

\newcommand{\Sem}{\mathcal{S}}

\newcommand{\PSPACE}{\textnormal{\textsf{PSPACE}}\xspace}

\newcommand{\car}{\mathbf{car}}

\newcommand{\diamTransf}{\mathbf{max\mbox{-}diam}}

\newcommand{\comment}[1]{}

\title{Careful synchronisation and the diameter \\ of transformation semigroups with few generators}
\author{Andrew Ryzhikov}
\date{University of Warsaw, Poland \\ {ryzhikov.andrew@gmail.com}}
\begin{document}

\maketitle              
\begin{abstract}
A word is called carefully synchronising for a partial deterministic finite semi-automaton if it maps all states to the same state. Equivalently, it is a composition of partial transformations equal to a constant total transformation. There is a sequence of several papers providing stronger and stronger lower bounds on the length of shortest carefully synchronising words for $n$-state partial DFAs over small alphabets. 
It resulted in the lower bounds of $\Omega(\frac{2^{n/3}}{n\sqrt{n}})$ and $\Omega(\frac{4^{n/5}}{n})$ for alphabets of two and three letters respectively, obtained by de Bondt, Don, and Zantema.
Using a significantly simpler construction, we improve these lower bounds to $2^{(n - 4)/3}$ and $4^{(n - 4)/5}$ respectively.

We then consider a tightly related question of the diameter of a partial DFA, which is the smallest $\ell \ge 0$ such that words of length at most $\ell$ express all the transformations induced by words in this DFA. We show that an alphabet of large enough constant size already asymptotically matches the upper bound on the diameter for arbitrary alphabet size, extending the construction of Panteleev that requires an alphabet of size exponential in the number of states. We then discuss an application to the diameter of finite semigroups of nonnegative matrices, and some open problems.

\vspace{0.15cm}
\end{abstract}
\section{Introduction}
A partial deterministic finite (semi-)automaton (partial DFA) $\Aa = (Q, \Sigma, \delta)$ consists of a finite set~$Q$ of states, a finite alphabet $\Sigma$, and a partial transition function $\delta \colon Q \times \Sigma \rightharpoonup Q$. The transition function can be homomorphically extended to $Q \times \Sigma^* \rightharpoonup Q$, which we also denote as $\delta$. When $\Aa$ is clear from the context, for $w \in \Sigma^*$ and $q \in Q$, we denote $\delta(q, w)$ as $q \cdot w$. If $q \cdot w$ is undefined, we write $q \cdot w = \perp$ and we say that $w$ \emph{kills}~$q$.
A word $w$ is called \emph{carefully synchronising} for $\Aa$ if there exists a state $q$ such that $p \cdot w = q$ for all $p \in Q$. In particular, a carefully synchronising does not kill any states. Intuitively, a carefully synchronising word can be thought of as a sequence of instructions bringing an automaton to a fixed state regardless of its current state, while avoiding instructions that somehow ``break'' the automaton (that is, that are not allowed when the automaton is in a particular state).

For total DFAs (that is, DFAs whose transition function is total), carefully synchronising words are known simply as synchronising words, and are the subject of the famous \v{C}ern\'{y} conjecture~\cite{Volkov2008,Kari2021}. Deciding if a total DFA admits a synchronising word can de done in polynomial time, and the length of a shortest such word is polynomial in the number of states~\cite{Volkov2008,Kari2021}. In contrast, deciding the existence of a carefully synchronising word is \PSPACE-complete~\cite{Martyugin2014}, and shortest such words can have exponential length in the number of states, as we discuss below.

Partial DFA $\Aa$ defines a natural homomorphism from the free semigroup of all words over $\Sigma$ to the semigroup of partial transformations on $Q$, assigning to each word $w$ its action $f$ on $Q$. Hence, we say that $w$ \emph{expresses} a partial transformation $f$ in $\Aa$, and the \emph{depth} of $f$ is the length of a shortest word expressing it.
The diameter of $\Aa$ is the maximum of the depths of transformations on $Q$ that are expressed in $\Aa$ by some word. This coincides with the notion of the diameter of the corresponding semigroup of partial transformations (with respect to a set of generators)~\cite{Panteleev2015}.
We denote by $\diamTransf(n, m)$ the maximum of the diameters of partial DFAs with~$n$ states and~$m$ letters. The diameter can be seen as the minimum length of a word serving as a natural witness for the membership problem, a fundamental computational problem for algebraic structures. Given an algebraic structure defined by a set of generators, this problem asks if a given element belongs to this structure.

\subsection*{Bounds on shortest carefully synchronising words} 

Denote by $\car(n, m)$ the maximum length of a shortest carefully synchronising word for partial DFAs with $n$ states and $m$ letters that admit a carefully synchronising word.
For the case where the alphabet is large enough it is well understood. Namely, $\car(n, m) \le 3^{(1 + o(1))\frac{n}{3}}$ for all $m$~\cite{Rystsov1980}, and this bound is asymptotically tight when $m$ grows linearly in $n$~\cite{Rystsov1980,Martyugin2010,Ryzhikov2018,Ruszil2022}. The upper bound is also partially rediscovered in~\cite{Gazdag2009}. For constant $m$, the situation is more complicated. 
In~\cite{Martyugin2008} (see also the last section
of~\cite{deBondt2018}), a superpolynomial in $n$ lower bound on $\car(n, 2)$ is provided, by using a construction with disjoint cycles of coprime length. A stronger almost-exponential lower bound on the same value is obtained in \cite{Martyugin2013}, which is improved to the exponential lower bound of ~$\Omega(2^{n/26})$  in~\cite{Vorel2016}. Finally, the best known lower bounds up to date are obtained in~\cite{Bondt2019} by an elegant but rather complicated construction. Let us discuss it in more detail.

In~\cite{Bondt2019}, it is shown that $\car(n, 2) = \Omega(\frac{2^{n/3}}{n\sqrt{n}})$ and $\car(n, 3) = \Omega(\frac{4^{n/5}}{n})$.
For a partial DFA $\Aa = (Q, \Sigma, \delta)$ and a word $w \in \Sigma^*$, we refer to the set of $Q \cdot w = \{q \cdot w \mid q \in Q\}$ as \emph{the set of active states after applying $w$}. This provides a convenient way of speaking about the images of $Q$ when a word is applied letter by letter.
The approach of~\cite{Bondt2019} is to encode a configuration of some rewriting system by active states in a partial DFA, so that the applied words that do not kill any states simulate rewriting rules. It is then shown that there exist two particular configurations such that going from the first one to the second one always requires a large number of steps, and that each carefully synchronising word corresponds to such a rewriting.
This approach poses two challenges for proving lower bounds and analysing the resulting DFAs. Firstly, the used rewriting system must be simple enough so that it is possible to simulate it by a DFA with few letters and without blowing up the number of states. Secondly, this rewriting system must be expressive enough to admit a long shortest sequence of rewritings. As a result, the construction in~\cite{Bondt2019} requires a rather intricate~analysis.

\subsection*{Bounds on the diameter of transformation semigroups}

By definition, for every $n$ and $m$ we have $\car(n, m) \le \diamTransf(n, m)$.
Let us also observe that, for estimating the asymptotics of $\diamTransf(n, m)$, it does not matter if one considers partial or total DFAs. Indeed, by adding a fresh state and sending all undefined transitions to it, we get a total DFA with the same diameter and only one more state.

In~\cite{Panteleev2015}, it is shown that $\diamTransf(n, m) \le 2^n e^{(1 + o(1))\sqrt{\frac{n}{2} \log n}}$ for every $n$ and $m$.
The key observation in the proof of this result is that the diameter of each $\mathcal{R}$-class can be bounded by the product of two values: the number of different images of transformations in this $\mathcal{R}$-class, and the diameter of a permutation group associated to it. This permutation group is in fact the generalised Sch\"{u}tzenberger group of an $\mathcal{R}$-class defined in \cite{Linton1998}, and Theorem 2.3 there implies a variant of this property. We refer to~\cite{Linton1998,Fleischer2019} for the mentioned definitions from semigroup theory.

In~\cite[Lemma 8]{Panteleev2015}, a construction showing that $\diamTransf(n, m) \ge 2^n e^{(1 + o(1))\sqrt{\frac{n}{2} \log n}}$ when~$m$ grows exponentially in $n$ is given, asymptotically matching the upper bound for all $m$.
Virtually the same construction is used in~\cite{Fleischer2019} to show a tight exponential lower bound on the maximum length of a chain of $\mathcal{R}$-classes in a transformation semigroup. This construction is then extended in the same paper to show that for a large enough constant-size alphabet the lower bound remains asymptotically the same.
The results of~\cite{Fleischer2019} do not immediately imply any lower bounds on $\diamTransf(n, m)$, since, intuitively, the diameter is related to the maximum length of a shortest chain, and not just the maximum length of any chain of $\mathcal{R}$-classes.
However, by using an idea similar to~\cite{Fleischer2019}, we get a tight lower bound on $\diamTransf(n, m)$ for large enough constant $m$. Moreover, the techniques developed in this paper allow to simplify and streamline this idea compared to its implementation in~\cite{Fleischer2019}.

\subsection*{Our contributions}

We provide a simple construction that improves the best known lower bounds on $\car(n, 3)$ (\Cref{thm:lb-car-ternary}) and $\car(n, 2)$ (\Cref{thm:lb-car-binary}). It is based on simulating decrements of a $4$-ary counter, and admits a simple analysis of both its behaviour and of the properties of the constructed DFA. The construction in the proof of \Cref{thm:lb-car-ternary} also implies a strong lower bound on $\diamTransf(n, 2)$ (\Cref{col-diam-bin}). We then use the developed techniques to show a simple lower bound for $\diamTransf(n, m)$ for a large enough constant $m$ (\Cref{thm:diam-lb-const}).  This lower bound asymptotically matches the upper bound for all~$m$. Finally, in \Cref{sec-conclusions} we discuss applications to matrix semigroups and mention several open problems.

\label{sec-intro}





\section{Careful synchronisation of partial DFAs}\label{sec-car}

\subsection{Linear-size alphabet}

We start by recalling and presenting in a more explicit way a construction from \cite[Corollary 3.9]{Ryzhikov2018} that simulates a decrementing ternary counter by a partial DFA. It will provide a basic intuition for the more involved constructions in this paper.

\begin{theorem}[\cite{Rystsov1980,Martyugin2010,Ryzhikov2018}]\label{thm:counter-car-lin}
    For every $n$ divisible by $3$, $\car(n, \frac{n}{3} + 2) \ge 3^{n/3} + 1$.
\end{theorem}

The idea of the proof is to construct a partial DFA $\Aa = (Q, \Sigma, \delta)$ with $n = 3k$ states, such that, when a carefully synchronising word is applied to it, the set of active states represents the ternary encoding of a number with $k$ digits. The transitions are defined in such a way that applying a letter that does not kill any active states decreases this number by one.
More precisely, for every digit of the representation, we will have three separate states, and exactly one of them will be active at a time, encoding the digit in the corresponding position. Careful synchronisation will be possible only when the encoded number goes all the way from $3^{k} - 1$ to $0$.

Formally, define the set $Q$ of states to be $Z \cup N \cup T$, where 
\[Z = \{z_1, \ldots, z_k\}, N = \{n_1, \ldots, n_k\}, T = \{t_1, \ldots, t_k\}.\]
Here, $Z$ stands for \underline{z}ero, $N$ for o\underline{n}e and $T$ for \underline{t}wo. We introduce a letter $c$, which is the only letter that does not kill any states, so it has to be applied first. We define
\[
    t_i \cdot c = n_i \cdot c = z_i \cdot c = t_i \mbox{ for all } 1 \le i \le k.
\]
After the first application of $c$, for each $i$, exactly one of $z_i, n_i, t_i$ is active. For the current set of active states, we define the ternary representation of the encoded number as $x_1 \cdots x_k$, where $x_i = 0$ if $z_i$ is active, $x_i = 1$ if $n_i$ is active, and $x_i = 2$ if $t_i$ is active. We define $k$ letters $a_1, \ldots, a_k$ so that letter $a_i$ can be used only if $x_{i + 1}, \ldots, x_k$ are all zeroes and $x_i$ is non-zero, by making $a_i$ kill an active state in all other situations. If $a_i$ does not kill any active states, its application decreases the encoded number by one. For example, for $i = 2$ we go from $1200$ to $1122$. Formally, for all $1 \le i \le k$,
\[
    t_j \cdot a_i = 
        \begin{cases}
         t_j & \mbox{if } j < i; \\
        n_j & \mbox{if } j = i; \\ 
        \perp & \mbox{if } j > i;
        \end{cases}\qquad
    n_j \cdot a_i = 
        \begin{cases}
         n_j & \mbox{if } j < i; \\
        z_j & \mbox{if } j = i; \\ 
        \perp & \mbox{if } j > i;
        \end{cases}
        \qquad
    z_j \cdot a_i = 
        \begin{cases}
         z_j & \mbox{if } j < i; \\
        \perp & \mbox{if } j = i; \\ 
        t_j & \mbox{if } j > i.
        \end{cases}
\]
Finally, we add a letter $y$ that allows to obtain a carefully synchronising word for $\Aa$ once the value of the encoded number reaches zero, by mapping all states in $Z$ to $z_k$ and killing all other states.
It is straightforward to see that $\Aa$ behaves exactly as described above, namely, after the first application of~$c$, the set of active states encodes a number, and this number is decremented by one whenever a letter that does not kill any active states is applied. To achieve careful synchronisation, this number must reach zero, which is the only situation where $y$ can be applied. Hence, the constructed DFA provides the lower bound from the statement of the theorem.

An illustration of the construction for $k = 4$ is provided below. The transitions by letter~$a_4$ are represented by solid arrows, by $a_2$ by dashed arrows, and by $y$ by dashdotted arrows. The transitions by $c$, $a_1$ and $a_3$ are omitted for the sake of clarity.

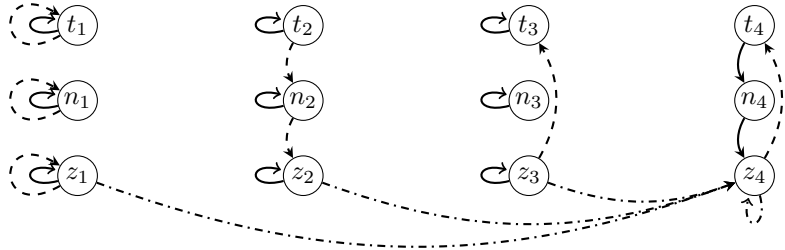
\begin{center}
\begin{tikzpicture} [node distance = 2cm]
\tikzset{every state/.style={inner sep=1pt,minimum size=1.5em}}

\node [state] at (0, 0) (z0) {$z_1$};
\node [state] at (3, 0) (z1) {$z_2$};
\node [state] at (6, 0) (z2) {$z_3$};
\node [state] at (9, 0) (z3) {$z_4$};

\node [state] at (0, 1) (n0) {$n_1$};
\node [state] at (3, 1) (n1) {$n_2$};
\node [state] at (6, 1) (n2) {$n_3$};
\node [state] at (9, 1) (n3) {$n_4$};

\node [state] at (0, 2) (t0) {$t_1$};
\node [state] at (3, 2) (t1) {$t_2$};
\node [state] at (6, 2) (t2) {$t_3$};
\node [state] at (9, 2) (t3) {$t_4$};

\path [-stealth, thick]

(t3) edge [bend right=30] node[left] {} (n3)
(n3) edge [bend right=30] node[left] {} (z3)
(z3) edge [bend right=30, dashed] node[right] {} (t3)

(z2) edge [bend right=30, dashed] node[right] {} (t2)

(t1) edge [bend right=30, dashed] node[left] {} (n1)
(n1) edge [bend right=30, dashed] node[left] {} (z1)


(z0) edge [bend right=20, dashdotted] node[right] {} (z3)
(z1) edge [bend right=20, dashdotted] node[right] {} (z3)
(z2) edge [bend right=20, dashdotted] node[right] {} (z3)
(z3) edge [loop below, dashdotted] node[right] {} (z3)

(t0) edge [loop left] node[left] {} (t0)
(n0) edge [loop left] node[left] {} (n0)
(z0) edge [loop left] node[left] {} (z0)
(t0) edge [loop left,min distance=10mm,in=150,out=-150, dashed] node[above] {} (t0)
(n0) edge [loop left,min distance=10mm,in=150,out=-150, dashed] node[above] {} (n0)
(z0) edge [loop left,min distance=10mm,in=150,out=-150, dashed] node[above] {} (z0)

(t1) edge [loop left] node[left] {} (t1)
(n1) edge [loop left] node[left] {} (n1)
(z1) edge [loop left] node[left] {} (z1)

(t2) edge [loop left] node[left] {} (t2)
(n2) edge [loop left] node[left] {} (n2)
(z2) edge [loop left] node[left] {} (z2)
;

\end{tikzpicture}
\end{center}

For $k = 4$, the unique shortest carefully synchronising word is
\[c (a_4 a_4) (a_3 a_4 a_4)^2 (a_2 (a_4 a_4) (a_3 a_4 a_4)^2)^2 a_1 (a_4 a_4) (a_3 a_4 a_4)^2 (a_2 (a_4 a_4) (a_3 a_4 a_4)^2)^2y.\]

\subsection{Ternary alphabet}\label{subsec:car-ternary}

We now extend the idea of the previous subsection to the case of an alphabet of constant size. Our approach is somewhat similar to the ideas in~\cite{Ryzhikov2024RP}, with several important differences. Namely, the approach in~\cite{Ryzhikov2024RP} relies on nondeterminism (the fact that a state can be mapped to multiple states by the same letter) and only allows to simulate a binary counter, since active states there correspond to ones, and inactive states to zeros. Showing how to get good enough lower bounds by simulating a counter in an arbitrary fixed base by a DFA is the main technical contribution of this paper. 

\begin{theorem}\label{thm:lb-car-ternary}
    For every $n \ge 9$, $n = 5k + 4$ with $k \in \mathbb{N}$, we have $\car(n, 3) \ge 4^{(n - 4)/5}$.
\end{theorem}

To slightly simplify the presentation, we construct a partial DFA $\Aa = (Q, \{c, d, r\}, \delta)$ with $n \ge 7$ states, $n = 4k + 3$, simulating a ternary counter. A partial DFA simulating a $4$-ary counter that provides the lower bound stated in the theorem is constructed in the same way. We discuss the choice of the optimal base after we describe the construction.

Similarly to the previous subsection, we define the set $Q$ of states as $Z \cup N \cup T \cup L \cup P$,
where 
\[Z = \{z_1, \ldots, z_k\}, N = \{n_1, \ldots, n_k\}, T = \{t_1, \ldots, t_k\}, L = \{\ell_0, \ell_1, \ell_2\}, P = \{p_1, \ldots, p_k\}.\]
States in $P$ will be used to store intermediate information while a word decrementing the counter by one is applied, and states in $L$ will be used to control which letter can be applied at every moment.

We define~$c$ to be the only letter that that does not kill any states, so it has to be applied first. After that, the set of active states becomes $T$. Formally, we define
\[
    t_i \cdot c = n_i \cdot c = z_i \cdot c = t_i \mbox{ for all } 1 \le i \le k;
\]
\[
    p_i \cdot c = t_{i + 1} \mbox{ for all } 1 \le i < k; \quad \ell_0 \cdot c = \ell_1 \cdot c = \ell_2 \cdot c = p_k \cdot c = t_1.
\]
Applying~$c$ afterwards will reset the counter to the largest value, so there will be no reason to apply it.

The counting process is performed as follows. At each step, for each $1 \le i \le k$, precisely one of $z_i, n_i, t_i$ is active, and the ternary encoding of the current value is defined as $x_1 \cdots x_k$, where $x_i = 0$ if $z_i$ is active, $x_i = 1$ if $n_i$ is active, and $x_i = 2$ if $t_i$ is active. Decrementing it by one is done in two phases. First, we remove zeroes at the end of the encoding by shifting it to the right (and mapping the corresponding active states to~$P$). To do that, we define letter~$d$.
As soon as the least significant digit in the encoding is non-zero, its corresponding active state is instead mapped by~$d$ to a state in $\{\ell_0, \ell_1\}$, and applying~$d$ is no longer allowed since it kills all states in $L$. Moreover, when an active state in mapped to a state in $\{\ell_0, \ell_1\}$, the corresponding digit is, intuitively, decremented by one. 
Formally, the action of~$d$ is defined as follows:
\[
    t_i \cdot d = 
        \begin{cases}
         t_{i + 1} & \mbox{if } 1 \le i < k; \\
        \ell_1 & \mbox{if } i = k; \end{cases}\qquad
    n_i \cdot d = 
        \begin{cases}
         n_{i + 1} & \mbox{if } 0 \le i < k; \\
        \ell_0 & \mbox{if } i = k; \end{cases}
\]
\[
    z_i \cdot d = 
        \begin{cases}
         z_{i + 1} & \mbox{if } 1 \le i < k; \\
        p_1 & \mbox{if } i = k; \end{cases}\qquad
    p_i \cdot d = 
        \begin{cases}
         p_{i + 1} & \mbox{if } 1 \le i < k; \\
        p_k & \mbox{if } i = k. \end{cases}
\]

Letter $d$ also kills all states in $L$. The illustration for $k = 5$ is provided below:

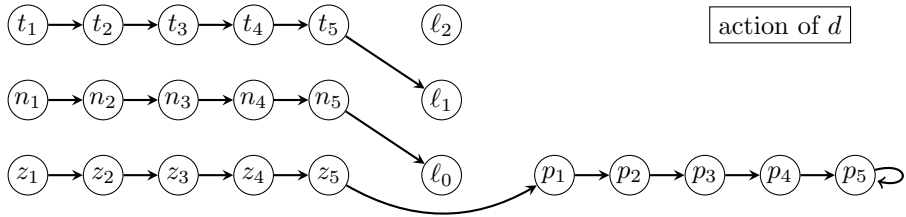
\begin{center}
\begin{tikzpicture} [node distance = 2cm]
\tikzset{every state/.style={inner sep=1pt,minimum size=1.5em}}

\node[draw] at (10,2) {action of $d$};

\node [state] at (0, 0) (z0) {$z_1$};
\node [state] at (1, 0) (z1) {$z_2$};
\node [state] at (2, 0) (z2) {$z_3$};
\node [state] at (3, 0) (z3) {$z_4$};
\node [state] at (4, 0) (z4) {$z_5$};

\node [state] at (0, 1) (n0) {$n_1$};
\node [state] at (1, 1) (n1) {$n_2$};
\node [state] at (2, 1) (n2) {$n_3$};
\node [state] at (3, 1) (n3) {$n_4$};
\node [state] at (4, 1) (n4) {$n_5$};

\node [state] at (0, 2) (t0) {$t_1$};
\node [state] at (1, 2) (t1) {$t_2$};
\node [state] at (2, 2) (t2) {$t_3$};
\node [state] at (3, 2) (t3) {$t_4$};
\node [state] at (4, 2) (t4) {$t_5$};

\node [state] at (7, 0) (q0) {$p_1$};
\node [state] at (8, 0) (q1) {$p_2$};
\node [state] at (9, 0) (q2) {$p_3$};
\node [state] at (10, 0) (q3) {$p_4$};
\node [state] at (11, 0) (q4) {$p_5$};

\node [state] at (5.5, 0) (l0) {$\ell_0$};
\node [state] at (5.5, 1) (l1) {$\ell_1$};
\node [state] at (5.5, 2) (lb) {$\ell_2$};

\path [-stealth, thick]

(z0) edge [] node[above] {} (z1)
(z1) edge [] node[above] {} (z2)
(z2) edge [] node[above] {} (z3)
(z3) edge [] node[above] {} (z4)
(z4) edge [bend right=30] node[above] {} (q0)

(n0) edge [] node[above] {} (n1)
(n1) edge [] node[above] {} (n2)
(n2) edge [] node[above] {} (n3)
(n3) edge [] node[above] {} (n4)
(n4) edge [] node[above] {} (l0)

(t0) edge [] node[above] {} (t1)
(t1) edge [] node[above] {} (t2)
(t2) edge [] node[above] {} (t3)
(t3) edge [] node[above] {} (t4)
(t4) edge [] node[above] {} (l1)

(q0) edge [] node[above] {} (q1)
(q1) edge [] node[above] {} (q2)
(q2) edge [] node[above] {} (q3)
(q3) edge [] node[above] {} (q4)

(q4) edge [loop right] node[above] {} (q4)

;

\end{tikzpicture}
\end{center}

As an example, if the currently encoded number is $10200$, letter $d$ can be applied three times, and the set of active states after that is $\{n_4, z_5, \ell_1, p_1, p_2\}$.

The second phase consists in shifting the number back to the left and simultaneously replacing all shifted zeros at the end with twos. To do that, we introduce letter $r$. Once $r$ is applied at least once, state $\ell_2$ becomes active and stays active until only states in $Z \cup N \cup T$ are active. Hence, we have to shift the encoding all the way back to the left before we can proceed to the next decrementing with~$d$, since $d$ kills $\ell_2$.
Formally, we define 
\[
    t_i \cdot r = 
        \begin{cases}
        \perp & \mbox{if } i = 1; \\
         t_{i - 1} & \mbox{if } 1 < i \le k;
         \end{cases}\qquad
    n_i \cdot r = 
        \begin{cases}
        \perp & \mbox{if } i = 1;\\
         n_{i - 1} & \mbox{if } 1 < i \le k;
         \end{cases}
\]
\[
    z_i \cdot r = 
        \begin{cases}
        \perp & \mbox{if } i = 1; \\
         z_{i - 1} & \mbox{if } 1 < i \le k;
         \end{cases}\qquad
    p_i \cdot r = 
        \begin{cases}
        \perp & \mbox{if } i = 0; \\
        \ell_2 & \mbox{if } i = 1; \\
         p_{i - 1} & \mbox{if } 2 \le i \le k.
         \end{cases}
\]

We also define $\ell_0 \cdot r = z_k$, $\ell_1 \cdot r = n_k$, $\ell_2 \cdot r = t_k$. The illustration for $k = 5$ is provided below:

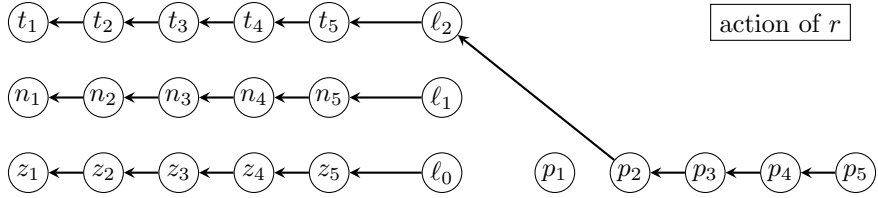
\begin{center}
\begin{tikzpicture} [node distance = 2cm]
\tikzset{every state/.style={inner sep=1pt,minimum size=1.5em}}

\node[draw] at (10, 2) {action of $r$};

\node [state] at (0, 0) (z0) {$z_1$};
\node [state] at (1, 0) (z1) {$z_2$};
\node [state] at (2, 0) (z2) {$z_3$};
\node [state] at (3, 0) (z3) {$z_4$};
\node [state] at (4, 0) (z4) {$z_5$};

\node [state] at (0, 1) (n0) {$n_1$};
\node [state] at (1, 1) (n1) {$n_2$};
\node [state] at (2, 1) (n2) {$n_3$};
\node [state] at (3, 1) (n3) {$n_4$};
\node [state] at (4, 1) (n4) {$n_5$};

\node [state] at (0, 2) (t0) {$t_1$};
\node [state] at (1, 2) (t1) {$t_2$};
\node [state] at (2, 2) (t2) {$t_3$};
\node [state] at (3, 2) (t3) {$t_4$};
\node [state] at (4, 2) (t4) {$t_5$};

\node [state] at (7, 0) (q0) {$p_1$};
\node [state] at (8, 0) (q1) {$p_2$};
\node [state] at (9, 0) (q2) {$p_3$};
\node [state] at (10, 0) (q3) {$p_4$};
\node [state] at (11, 0) (q4) {$p_5$};

\node [state] at (5.5, 0) (l0) {$\ell_0$};
\node [state] at (5.5, 1) (l1) {$\ell_1$};
\node [state] at (5.5, 2) (lb) {$\ell_2$};

\path [-stealth, thick]

(z1) edge [] node[above] {} (z0)
(z2) edge [] node[above] {} (z1)
(z3) edge [] node[above] {} (z2)
(z4) edge [] node[above] {} (z3)
(q1) edge [] node[above] {} (lb)
(lb) edge [] node[above] {} (t4)

(n1) edge [] node[above] {} (n0)
(n2) edge [] node[above] {} (n1)
(n3) edge [] node[above] {} (n2)
(n4) edge [] node[above] {} (n3)
(l1) edge [] node[above] {} (n4)

(t1) edge [] node[above] {} (t0)
(t2) edge [] node[above] {} (t1)
(t3) edge [] node[above] {} (t2)
(t4) edge [] node[above] {} (t3)
(l0) edge [] node[above] {} (z4)

(q2) edge [] node[above] {} (q1)
(q3) edge [] node[above] {} (q2)
(q4) edge [] node[above] {} (q3)

;

\end{tikzpicture}
\end{center}

Observe that, when constructing a carefully synchronising word, at every moment there is only one letter that can be applied, assuming that letter~$c$ is only applied once. Since~$c$ makes the set $T$ active, it just resets the encoded value to the maximum one, and hence there is no reason to apply it more than once. After~$d$ is applied, exactly one state in $L \cup \{p_1\}$ becomes active.
Since~$r$ kills~$p_1$, $d$ has to be applied until~$p_1$ is not active, that is, until a state in $\{\ell_0, \ell_1\}$ becomes active. Then~$r$ must be applied, and~$\ell_2$ becomes active. Thus,~$r$ has to be applied until only states in $Z \cup N \cup T$ are active, since $d$ kills~$\ell_2$. Further shifts to the left by~$d$ are not possible, since $d$ kills $z_1, n_1, t_1$. 

The only way to carefully synchronise~$\Aa$ is thus to make the encoded value zero, which is the only way to make only states in $P$ active. By repeatedly applying~$d$ afterwards, this set can finally be synchronised. Thus, the length of a shortest synchronising word is at least $3^{(n - 3)/4}$.

By simulating a $b$-ary counter, we obtain that $\car(n, 3) \ge b^{(n - b)/(b + 1)}$. It is thus easy to see that for positive integer values of~$b$, the largest lower bound is obtained when $b = 4$, which concludes the proof of the theorem.

\subsection{Binary alphabet}\label{subsec:car-binary}

The adaptation of the construction from the previous subsection to the case of a binary alphabet is fairly simple. The idea is that we want to get rid of letter~$c$, and for that we need to make~$r$ defined for all states without creating any shortcuts for counting. Namely, excessive shifts to the left should not decrement the encoded value or merge any active states. We achieve that by introducing~$k$ fresh states which, when used, reset the encoded number to the maximum value. The result that we prove in this subsection is as follows.

\begin{theorem}\label{thm:lb-car-binary}
    For every $n \ge 10$, $n = 6k + 4$ with $k \in \mathbb{N}$, we have $\car(n, 2) \ge 2^{(n - 4)/3}$.
\end{theorem}

As in the previous subsection, the optimal base for the counter is $4$, but for the sake of simplicity we continue with a ternary counter, extending the construction from the previous subsection.
We introduce a fresh set $B = \{b_1, \ldots, b_k\}$ of states, and define
\[b_i \cdot d = b_{i + 1} \mbox{ for all } 1 \le i < k \mbox{ and } b_k \cdot d = t_1.\]


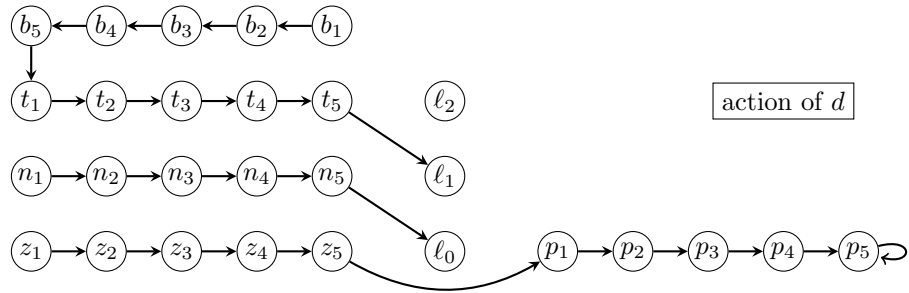
\begin{center}
\begin{tikzpicture} [node distance = 2cm]
\tikzset{every state/.style={inner sep=1pt,minimum size=1.5em}}

\node[draw] at (10, 2) {action of $d$};

\node [state] at (4, 3) (b0) {$b_1$};
\node [state] at (3, 3) (b1) {$b_2$};
\node [state] at (2, 3) (b2) {$b_3$};
\node [state] at (1, 3) (b3) {$b_4$};
\node [state] at (0, 3) (b4) {$b_5$};

\node [state] at (0, 0) (z0) {$z_1$};
\node [state] at (1, 0) (z1) {$z_2$};
\node [state] at (2, 0) (z2) {$z_3$};
\node [state] at (3, 0) (z3) {$z_4$};
\node [state] at (4, 0) (z4) {$z_5$};

\node [state] at (0, 1) (n0) {$n_1$};
\node [state] at (1, 1) (n1) {$n_2$};
\node [state] at (2, 1) (n2) {$n_3$};
\node [state] at (3, 1) (n3) {$n_4$};
\node [state] at (4, 1) (n4) {$n_5$};

\node [state] at (0, 2) (t0) {$t_1$};
\node [state] at (1, 2) (t1) {$t_2$};
\node [state] at (2, 2) (t2) {$t_3$};
\node [state] at (3, 2) (t3) {$t_4$};
\node [state] at (4, 2) (t4) {$t_5$};

\node [state] at (7, 0) (q0) {$p_1$};
\node [state] at (8, 0) (q1) {$p_2$};
\node [state] at (9, 0) (q2) {$p_3$};
\node [state] at (10, 0) (q3) {$p_4$};
\node [state] at (11, 0) (q4) {$p_5$};

\node [state] at (5.5, 0) (l0) {$\ell_0$};
\node [state] at (5.5, 1) (l1) {$\ell_1$};
\node [state] at (5.5, 2) (lb) {$\ell_2$};

\path [-stealth, thick]

(b0) edge [] node[above] {} (b1)
(b1) edge [] node[above] {} (b2)
(b2) edge [] node[above] {} (b3)
(b3) edge [] node[above] {} (b4)
(b4) edge [] node[above] {} (t0)

(z0) edge [] node[above] {} (z1)
(z1) edge [] node[above] {} (z2)
(z2) edge [] node[above] {} (z3)
(z3) edge [] node[above] {} (z4)
(z4) edge [bend right=30] node[above] {} (q0)

(n0) edge [] node[above] {} (n1)
(n1) edge [] node[above] {} (n2)
(n2) edge [] node[above] {} (n3)
(n3) edge [] node[above] {} (n4)
(n4) edge [] node[above] {} (l0)

(t0) edge [] node[above] {} (t1)
(t1) edge [] node[above] {} (t2)
(t2) edge [] node[above] {} (t3)
(t3) edge [] node[above] {} (t4)
(t4) edge [] node[above] {} (l1)

(q0) edge [] node[above] {} (q1)
(q1) edge [] node[above] {} (q2)
(q2) edge [] node[above] {} (q3)
(q3) edge [] node[above] {} (q4)

(q4) edge [loop right] node[above] {} (q4)
(z0) edge [bend left=40,color=white] node[above] {} (b4)

;

\end{tikzpicture}
\end{center}

For the action of $r$, we define 

\[b_i \cdot r = b_{i - 1} \mbox{ for all } 2 \le i \le k \mbox{ and } b_1 \cdot r = b_k;\]
\[t_1 \cdot r = n_1 \cdot r = z_1 \cdot r = b_k; \quad q_1 \cdot r = t_k.\]
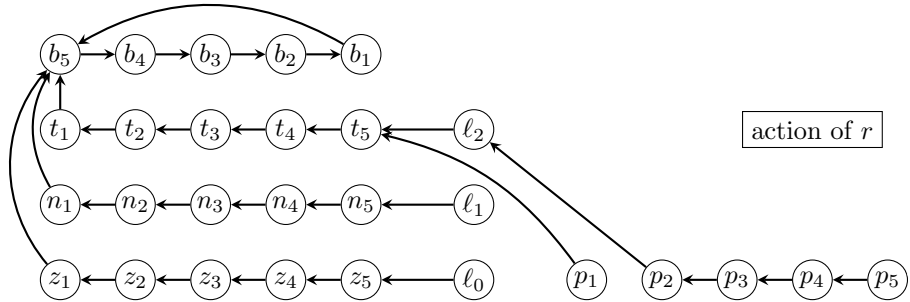
\begin{center}
\begin{tikzpicture} [node distance = 2cm]
\tikzset{every state/.style={inner sep=1pt,minimum size=1.5em}}

\node[draw] at (10, 2) {action of $r$};

\node [state] at (4, 3) (b0) {$b_1$};
\node [state] at (3, 3) (b1) {$b_2$};
\node [state] at (2, 3) (b2) {$b_3$};
\node [state] at (1, 3) (b3) {$b_4$};
\node [state] at (0, 3) (b4) {$b_5$};

\node [state] at (0, 0) (z0) {$z_1$};
\node [state] at (1, 0) (z1) {$z_2$};
\node [state] at (2, 0) (z2) {$z_3$};
\node [state] at (3, 0) (z3) {$z_4$};
\node [state] at (4, 0) (z4) {$z_5$};

\node [state] at (0, 1) (n0) {$n_1$};
\node [state] at (1, 1) (n1) {$n_2$};
\node [state] at (2, 1) (n2) {$n_3$};
\node [state] at (3, 1) (n3) {$n_4$};
\node [state] at (4, 1) (n4) {$n_5$};

\node [state] at (0, 2) (t0) {$t_1$};
\node [state] at (1, 2) (t1) {$t_2$};
\node [state] at (2, 2) (t2) {$t_3$};
\node [state] at (3, 2) (t3) {$t_4$};
\node [state] at (4, 2) (t4) {$t_5$};

\node [state] at (7, 0) (q0) {$p_1$};
\node [state] at (8, 0) (q1) {$p_2$};
\node [state] at (9, 0) (q2) {$p_3$};
\node [state] at (10, 0) (q3) {$p_4$};
\node [state] at (11, 0) (q4) {$p_5$};

\node [state] at (5.5, 0) (l0) {$\ell_0$};
\node [state] at (5.5, 1) (l1) {$\ell_1$};
\node [state] at (5.5, 2) (lb) {$\ell_2$};

\path [-stealth, thick]

(b0) edge [bend right=30] node[above] {} (b4)
(b1) edge [] node[above] {} (b0)
(b2) edge [] node[above] {} (b1)
(b3) edge [] node[above] {} (b2)
(b4) edge [] node[above] {} (b3)
(t0) edge [] node[above] {} (b4)

(z1) edge [] node[above] {} (z0)
(z2) edge [] node[above] {} (z1)
(z3) edge [] node[above] {} (z2)
(z4) edge [] node[above] {} (z3)
(z0) edge [bend left=40] node[above] {} (b4)

(q1) edge [] node[above] {} (lb)
(q0) edge [bend right=20] node[above] {} (t4)
(lb) edge [] node[above] {} (t4)

(n1) edge [] node[above] {} (n0)
(n2) edge [] node[above] {} (n1)
(n3) edge [] node[above] {} (n2)
(n4) edge [] node[above] {} (n3)
(l1) edge [] node[above] {} (n4)
(n0) edge [bend left=30] node[above] {} (b4)

(t1) edge [] node[above] {} (t0)
(t2) edge [] node[above] {} (t1)
(t3) edge [] node[above] {} (t2)
(t4) edge [] node[above] {} (t3)
(l0) edge [] node[above] {} (z4)

(q2) edge [] node[above] {} (q1)
(q3) edge [] node[above] {} (q2)
(q4) edge [] node[above] {} (q3)

(q4) edge [loop right, color=white] node[above] {} (q4)
;

\end{tikzpicture}
\end{center}

Note that in the very beginning, only letter~$r$ can be applied, and it has to be applied enough times so that only states in the set $B$ are active. Any further application of $r$ clearly does not change the set of active states, and $d$ has to be applied at least $k$ times, thus making the set $T$ active. Starting from there, we track the value encoded by the set of active states, in the same way as in the previous subsection. The only difference is that some active states can now be shifted more to the left, to $B$. However, to perform careful synchronisation, these active states have to be brought back to $Z \cup N \cup T$ afterwards. Shifting active states to $B$ and then back to $Z \cup N \cup T$ results in setting the corresponding digits of the encoding to $3$. Hence, by doing that the encoded number can only increase. The same reasoning applies to $p_1$: if it is active when $r$ is applied, the corresponding digits will be set to $3$ in the encoding. Hence, the same lower bound on the minimum length of a carefully synchronising word as in the previous subsection still applies, with the difference that now the constructed DFA has $k$ more states.

\section{The diameter of transformation semigroups}\label{sec-diameter}
The previous section can be seen as estimating the depth of a specific transformation in a partial DFA, namely a constant total transformation. We now turn our attention to estimating the depth of all transformations in the transition semigroup. While the ideas developed for careful synchronisation are still helpful, we are able to show stronger lower bounds.

Let us first consider the case of a total DFA with $n$ states and one letter. If this letter acts as a permutation on the state set, then the diameter of the DFA is clearly equal to the order of this permutation. The maximum order of a permutation on a set of $n$ elements is $e^{(1 + o(1))\sqrt{\frac{n}{2} \log n}}$ (see e.g.~\cite{Miller1987}). We will need such permutations in our construction, so for each $n$, denote by $\gamma_n$ an arbitrary permutation of maximum order among all permutations on $n$ elements, and by $g(n)$ its order. Intuitively, this permutation consists of many cycles with coprime lengths.
It is easy to see that in the case of a unary alphabet, a total DFA with a letter acting as a permutation has the largest possible diameter, hence $\diamTransf(n, 1) =  g(n)$. 

For the case of a binary alphabet, we can use the construction from \Cref{subsec:car-ternary}, to get a lower bound on  $\diamTransf(n, 2)$: remove letter~$c$ and observe that the depth of any transformation mapping all states in $T$ to $p_k$ is at least~$4^{(n - 4)/5}$. Hence, we get the following result.

\begin{corollary}\label{col-diam-bin}
    For every $n \ge 9$, $n = 5k + 4$ with $k \in \mathbb{N}$,  $\diamTransf(n, 2) \ge 4^{(n - 4)/5}$.
\end{corollary}

To the best of our knowledge, the only previously known lower bounds on $\diamTransf(n, 2)$ are provided by lower bounding $\car(n, 2)$, and are thus smaller.

The goal of this section is to prove that for some constant~$m$, the lower bound on $\diamTransf(n, m)$ asymptotically matches the upper bound for all $m$ proved in~\cite{Panteleev2015}. Our proof can be seen as a variation of the technique from~\cite{Fleischer2019}. The main contribution of this section is thus providing a simpler implementation of this technique and unifying the treatment of the diameter and the longest chain of $\mathcal{R}$-classes. It is interesting to remark that the construction providing an asymptotically tight lower bound on the longest chain of $\mathcal{R}$-classes provides an equally large lower bound on the diameter, but the converse does not hold: in the construction for the diameter all chains of $\mathcal{R}$-classes are short.

\begin{theorem}\label{thm:diam-lb-const}
    There exists a constant $m$ such that for all $n$, $\diamTransf(n, m) \ge 2^n e^{(1 + o(1))\sqrt{\frac{n}{2} \log n}}$.
\end{theorem}

\subparagraph*{Exponential-size alphabet construction.}
We start by recalling the construction from the proof of \cite[Lemma 8]{Panteleev2015}. For every $k \ge 1$, it provides a partial DFA~$\Aa = (Q, \Sigma, \delta)$ with $n = 2k$ states and $\binom{2k}{k} + 2$ letters, and a transformation $f$ such that the depth of $f$ in $\Aa$ is at least $\binom{2k}{k} \cdot g(k)$. Note that $\binom{2k}{k} \cdot g(k) \ge  2^n e^{(1 + o(1))\sqrt{\frac{n}{2} \log n}}$. Define $Q = \{1, \ldots, 2k\}$. For now, we require that the image of $Q$ under $f$ is $\{k + 1, \ldots, 2k\}$, and that $f$ is total. We will fully define $f$ later.

\textit{Initialisation.} First, we define a letter $c$ that maps $i$ and $i + k$ to $i$, for all $1 \le i \le k$. It will be the only letter that does not kill any states, so that it is the first letter in any word expressing $f$. After applying~$c$, the number of active states must always be $k$. We call a subset of $Q$ of size $k$ a \emph{$k$-subset}.

\textit{Visiting all $k$-subsets.} Fix an arbitrary ordering of all $k$-subsets such that $\{1, \ldots, k\}$ is the first one, and $\{k + 1, \ldots, 2k\}$ is the last one. For each $k$-subset $S$ except the last one in the ordering, define a separate letter that injectively maps states from $S$ to the next $k$-subset in the ordering, and kills all other states.
Clearly, in any word expressing~$f$, each letter has to occur at least once, hence the length of any such word is at least $\binom{2k}{k}$. 

\textit{Permuting elements in $k$-subsets.} So far, we only specified the images of $k$-subsets, and not of specific states. We can thus assume that the actions of all letters defined so far preserve the relative order of the states which they do not kill. That is, we assume that if $i < j$ for $i, j \in Q$, then $i \cdot x < j \cdot x$ if letter $x$ does not kill $i$ and $j$.

We now add another letter $y$ that kills all states in $\{1, \ldots, k\}$. It can thus only be applied when the set of active states is $\{k + 1, \ldots, 2k\}$. For each $i$, $1 \le i \le k$, we define $y$ to map $i + k$ to~$\gamma_k(i)$.
We require $f$ to map $i$ to $k + \gamma_k^{g(k) - 1}(i)$, where $\gamma_k^{g(k) - 1}$ is $g(k) - 1$ compositions of $\gamma_k$.
Since the action of each letter except $y$ does not change the relative order of active states, $y$ has to be applied at least $g(k) - 1$ times. After each application of $y$, the set of active states becomes the first $k$-subset in the ordering that we fixed, so a word of length at least $\binom{2k}{k}$ must be applied between every two application of $y$, and after the last application. Hence, the depth of $f$ is at least $\binom{2k}{k} \cdot (g(k) - 1) + \binom{2k}{k} = \binom{2k}{k} \cdot g(k)$. By construction, a word representing $f$ exists.

\subparagraph*{Overview of our construction.}
To obtain an analogous construction using an alphabet of constant size, we define the actions of the letters in such a way that every word $w$ expressing~$f$ can be represented as $w = w_1 w_2 \cdots w_\ell$, where for each $h$, $Q \cdot w_1 w_2 \cdots w_h$ is equal to the $h$th $k$-subset in the specific ordering of $k$-subsets that we fix. To guarantee that, we show how to define the same regular constraints on each~$w_h$. These constraints, intuitively, say that the current set of active states must be mapped by $w_h$ to the next $k$-subset in the ordering.

In more detail, we say that a word $u$ is \emph{$f$-extendable} if there exists a word $v$ such that $uv$ expresses~$f$. Observe that by adding a constant number of fresh states and fresh letters, we can ensure that all $f$-extendable words belong to a regular language recognised by a fixed DFA $\mathcal{B}$. We can assume that there is only one accepting state in $\mathcal{B}$ by adding a fresh letter and a fresh state. Add the states and the letters of $\mathcal{B}$ to the construction of $\Aa$ and require~$f$ to map the initial state of $\mathcal{B}$ to the accepting state. We can thus define a language constraining words~$w_h$, and then use its Kleene star, which is also a fixed regular language, to constrain words expressing~$f$.
In the sequel, we refer to constructing a single $w_h$ as one step in constructing a word expressing~$f$.

\subparagraph*{Lexicographically ordering $k$-subsets.} 
We fix the following lexicographic order on $k$-subsets. Let $S_1, S_2$ be two $k$-subsets, and let $b$ be the smallest element that belongs to exactly one of them. 
We define $S_1 < S_2$ if and only if $b$ belongs to~$S_1$. If we consider $k$-subsets as words over alphabet~$Q$ with elements written in the increasing order, this is the usual lexicographic order on words.

Let $S$ be a $k$-subset.
By definition, the next $k$-subset $S'$ in the lexicographic ordering is constructed as follows. 
Let $i$ be the largest element of $Q$ such that $i \in S$, $i + 1 \not \in S$ (if there is no such~$i$, $S$ is already lexicographically largest). Let $j$ be the largest element of $Q$ such that $j \not \in S$ and $S$ contains all elements larger than $j$. Then $S'$ is obtained from $S$ by removing $i$ and all  elements larger than $j$, and adding $i + 1, i + 2, \ldots, i + 2k - j + 1$ instead. For example, if $2k = 10$ and $S = \{3, 4, 8, 9, 10\}$, then $i = 4, j = 7$, and $S' = \{3, 5, 6, 7, 8\}$.

\subparagraph*{Guessing the values of $i$ and $j$.} 
To represent intermediate values, we extend the construction of a counter from \Cref{subsec:car-ternary}. For simplicity, we use counters in a binary base. Our extension has at most $3\log (2k + 1)$ fresh states and a constant number of fresh letters, and can store a number between $0$ and $2k$, decrement it by one, test equality to zero and to $k$ and reset the value back to $2k$. We assume that each counter starts from the value $2k$, by adjusting the action of letter~$c$ from the construction in \Cref{subsec:car-ternary} accordingly.
Decrementing the value by one is done by requiring that letter~$d$ is applied at least once. To make sure that enough left shifts by letter $r$ are performed after that, we introduce another letter that kills all states that we do not want to be active, and then require that this letter is applied each time after $d$ is applied. To test equality to zero or $k$, we do the same. Clearly, all these constraints define a regular language, and hence can be implemented as discussed above. We extend the definition of $f$ to require that no active state in any counter gadget must be killed.

We introduce two separate  counter gadgets $\mathbf{c_i}$ and $\mathbf{c_j}$ that will store the values of~$i$ and~$j$. We guess their values at the beginning of each step by resetting them to $2k$ and then, intuitively, nondeterministically decrementing them. More formally, we allow an arbitrary number of decrements, but we define the actions of the letters in such a way that getting to a wrong value results in a word that is no longer $f$-extendable. This is formally described below.

\subparagraph*{Verifying the guesses.} 
To verify the guessed values of $i$ and $j$, we need to check that states $i$ and  $j + 1\ldots, 2k$ are active, and states $i + 1, \ldots, j$ are not active.
To check that a state is not active, we can simply define a letter that kills this state, an require that this letter must be applied during a step.
Checking that a state is active is less straightforward. To check that all states in a set $S$ are active, we instead check that the number of inactive states in $Q \setminus S$ is $k$.

Introduce a letter $a$ that cyclically permutes the states in $Q$, that is, maps $h$ to $h - 1$ for $2 \le h \le 2k$ and maps $1$ to $2k$. 
Each step consists of performing $2k$ cyclic shifts, which can be ensured by introducing a separate counter storing the number of remaining shifts and decrementing it after each shift. The shifts subsequently move each state in $Q$ to state $1$ one by one.  
Define a fresh counter gadget $\mathbf{c_{act}}$ that will be counting the number of active states. After each cyclic shift by $a$, we decrement the values of $\mathbf{c_i}$ and~$\mathbf{c_j}$ by one to take into account the cyclic shift (if their values are still positive), and then proceed depending on their new values. 

\textit{Stage 1.} If the value of $\mathbf{c_i}$ is still greater than zero, we make a nondeterministic guess whether state~$1$ is currently active. If we guess that it is not, we require a letter that kills state~$1$ to be applied immediately. After each application of this letter, we decrement $\mathbf{c_{act}}$ by one. 

\textit{Stage 2.} Once the value of $\mathbf{c_i}$ becomes zero, we perform one cyclic shift to account for our assumption that state~$i$ was active in the beginning of the step. Then, until the value of $\mathbf{c_j}$ is zero, we perform a cyclic shift by~$a$ and decrement $\mathbf{c_j}$ by one. After each such operation, we require that a letter killing $1$ is immediately applied, thus verifying that the states $i + 1, \ldots, j$ were all inactive at the beginning of the step. Each time we decrement $\mathbf{c_{act}}$ by one.

\textit{Stage 3.} Finally, when $\mathbf{c_j}$ reaches zero, we apply~$a$ until the total number of cyclic shifts reaches~$2k$. We then check if the value of $\mathbf{c_{act}}$ is $k$, which holds true if and only if all our guesses were correct. If some our guess was incorrect, an active state from $\mathbf{c_{act}}$ will be killed, and the constructed word will no longer be $f$-extendable.

\subparagraph*{Going to the next $k$-subset.}
We now describe what happens at the end of each step. By storing in fresh counters the  values of $i$ and $j$ guessed in the beginning of the step, it is easy to construct a procedure similar to the verification procedure above to make the next $k$-subset in the lexicographic ordering active. Namely, apply $a$ until state $i$ is mapped to $1$ and require a fresh letter mapping state $1$ to $2$ and keeping all other states in their places to be applied immediately. Similarly, every active state in $j + 1, \ldots, 2k$ can be mapped to its new position by using a fixed number of fresh counters. 

It remains to define a letter $y$ in the same way as it is done in the case of an exponential-size alphabet above to obtain the proof of the theorem. 

\subparagraph*{Long chain of $\mathcal{R}$-classes.}
Let us remark that the existence of a chain of $\mathcal{R}$-classes of length $\ell$ (see~\cite{Fleischer2019} for a formal definition)  in a DFA is implied by the following property: there exist words $w, w_1, \ldots, w_\ell$, $w = w_1 w_2 \cdots w_\ell$, such that for each $1 \le h \le \ell$ and each word~$u$, the transformation expressed by $w_1 \cdots w_h u$ is not equal to any of the transformations expressed by $w_1 \cdots w_{h'}$ with $h' < h$. Hence, if we do not add letter $y$, and also add a regular constraint that $c$ can be applied only once, we get a DFA with a chain of $\mathcal{R}$-classes of length~$\binom{2k}{k}$, thus obtaining a simpler proof of \cite[Theorem~2]{Fleischer2019}. Indeed, as already discussed above, any word expressing $f$ makes a new $k$-subset in the lexicographic ordering active at the beginning of each step, and such a subset cannot become active again in the beginning of any subsequent step.


\section{Conclusions and open problems}\label{sec-conclusions}
We believe that the results of this paper together with~\cite{Ryzhikov2024RP} prove that the technique of simulating counters is a powerful and simple instrument for constructing lower bounds for extremal problems about automata, transformation semigroups and semigroups of nonnegative matrices. It is curious that our lower bounds for careful synchronisation asymptotically match the ones in~\cite{Bondt2019}, and our results provide a simple explanation of where the values $2^{\frac{2n}{5}} (= 4^{\frac{n}{5}})$ and $2^{\frac{n}{3}} (= 4^{\frac{n}{6}})$ in these lower bounds come from. Another example of such coincidence is the application of lower bounds on $\car(n, m)$ to lower bounds on the length of shortest mortal words in semigroups of nonnegative matrices described in \cite[Subsection 2.2]{Ryzhikov2024RP}. 
For $m = 2$, this application matches the lower bound in \cite[Theorem~6]{Ryzhikov2024RP}, thus providing an alternative construction for the proof of this theorem. The implied lower bound for $m = 3$ is however weaker. We conjecture that for the three mentioned instances where the lower bounds between different constructions match, these lower bounds are tight. 

It seems that our constructions for careful synchronisation cannot be extended any further: intuitively, there must be a subset of states that stores the information between decrements, and this subset must be bigger if the alphabet is binary. We hope however that this intuition might be insightful for proving upper bounds on $\car(n, 3)$ and $\car(n, 2)$, which is the main open problem left by this work. Another open problem is to establish a better upper bound on $\diamTransf(n, 2)$.

It would be interesting to see if some property of DFAs forbids simulating a counter in some formal sense. This may result in a class with polynomial upper bounds on $\car(n, m)$  and $\diamTransf(n, m)$. 
An advantage of our approach is that the structure of the constructed DFAs can be easily analysed, which in particular allows to refute candidates for such a class. For example, it is easy to see that the constructions in \Cref{subsec:car-ternary} and \Cref{subsec:car-binary} are strongly connected, that is, the underlying digraphs of the DFAs are strongly connected.

We remark that bounds on $\car(n, m)$ can be applied to studying primitive sets of matrices~\cite{Gerencser2018}. A more straightforward representation by matrices allows to show a rather strong lower bound on the diameter of finite semigroups of nonnegative integer matrices, which is defined analogously to the transformation case. 
Namely, consider a natural homomorphism from a semigroup of transformations on $n$ elements to a semigroup of $n \times n$ matrices with entries in $\{0, 1\}$. The results of \Cref{sec-diameter} then directly apply to the diameter of such matrix semigroups. This improves the result of~\cite[Theorem 2.2]{Weber1991}, which states that for every $n \ge 3$ there exists a set of $n \times n$ nonnegative integer matrices of cardinality $n$ generating a finite semigroup of diameter at least $2^{n - 2}$. It is worth mentioning that the proof of this result can be once again seen as simulating a binary counter, in the spirit of the description in the proof of~\cite[Theorem~4]{Ryzhikov2024RP}.





\subsubsection*{Acknowledgments} The author is grateful to Anton Varonka for useful discussions and comments on the manuscript. Andrew Ryzhikov is supported by Polish National Science Centre SONATA BIS-12 grant
number 2022/46/E/ST6/00230.

%
%
%
 \bibliographystyle{alpha}
 \bibliography{ref}
\end{document}